# Theoretical Model and Characteristics of Mitochondrial Thermogenesis

Jian-Sheng Kang[#]

The First Affiliated Hospital of Zhengzhou University, Zhengzhou, Henan, 450052, P. R. China

Text: 1697

# Correspondence should be addressed to J.-S. Kang, The First Affiliated Hospital of Zhengzhou University, #1 Jianshe East Road, Zhengzhou, Henan, 450052, P. R. China. E-mail: kjs@zzu.edu.cn

ORCID ID: 0000-0002-2603-9718

**Abstract**

Based on the first law of thermodynamics and the thermal diffusion equation, the deduced theoretical model of mitochondrial thermogenesis satisfies the Laplace equation and is a special case of the thermal diffusion equation. The model settles the long-standing question of the ability to increase cellular temperature by endogenous thermogenesis and explains the thermogenic characteristics of brown adipocytes. The model and calculations also suggest that the number of free available protons is the major limiting factor for endogenous thermogenesis and its speed.

**Highlights:**

- Mitochondrial thermogenesis is a special case of the thermal diffusion equation ($\nabla^2 T = 0$)

- The thermogenic rate is proportional to electrochemical potential energy dissipation

- The free proton number is the limiting factor for endogenous thermogenesis and its speed

- The theoretical model is generalized for all types of cells

Mitochondria are the main intracellular sites for thermogenesis, especially the mitochondria of brown adipocytes (BA), which have been targeted for therapy to reduce obesity. However, long-standing critique[1] and debate[2,3,4,5,6] exist on the ability to increase cellular temperature by endogenous thermogenesis, and a good theory of intracellular thermogenesis and temperature is lacking. In this work, based on the first law of thermodynamics and the thermal diffusion equation, the thermal physical model of a mitochondrion is deduced. We found that mitochondrial thermogenesis is a special case of the thermal diffusion equation, which satisfies the Laplace equation ($\nabla^2 T = 0$).

The cell is a membrane-enclosed grand canonical ensemble of systems that exchanges both heat and particles with its surroundings. We use the first law of thermodynamics to show that

$$U = -Q + W \quad (1)$$

where $U$ is the cellular internal energy, $Q$ is the heat dissipated to the surroundings, and $W$ is the work added to the system. In this equation, the negative sign means that heat flows out of the cell.

For a differential change, the relation in equation (1) is given by the following

$$dU = -dQ + dW \quad (2)$$

For thermogenesis of BA, a cell with such a small size has limited sources for energy extraction or delivery compared with the amplitudes of heat ($dQ$) and work ($dW$). Thus, it is acceptable to claim that the change in cellular internal energy ($dU$) can be neglected.

$$dU \approx 0 \quad (3)$$

Therefore, relation (2) is reduced to the following

$$dQ = dW \quad (4)$$

We can write the cellular work ($dW$) as the sum of various forms, such as kinetic energy and

potential energy:

$$dW = -pdV + Fdx + \sum_i \mu_i dN_i + \sum_j \varphi_j dq_j \quad (5)$$

where *pdV* is the work performed by the volume change (*dV*) under pressure (*p*), and *Fdx* is the mechanical energy used to move a distance (*dx*) under force (*F*). In addition to kinetic energy, potential energy contains chemical potential ($\mu$) and electric potential ($\varphi$) in the case of changing the numbers of particles (*dN*) or charges (*dq*).

Thermogenesis in BA is executed at the mitochondrial level. A single BA contains numerous mitochondria, which show minimal volume change and almost no motility in such a crowded space[7,8]. It is acceptable to consider that both *dV* and *dx* are equal to zero such that we can ignore the changes in kinetic energy and only consider the changes in potential energy. Thus, relation (5) is reduced to the following

$$dW = \sum_i \mu_i dN_i + \sum_j \varphi_j dq_j \quad (6)$$

A mitochondrion with a large negative membrane potential has the proton-motive force (*pmf*) for ATP synthase as well as motive forces (*mf*) for other particles, such as $Ca^{2+}$, among others. Thus, we can write relation (6) as follows

$$dW = pmf \cdot dH^+ + mf_{Ca^{2+}} \cdot dCa^{2+} + \cdots \quad (7)$$

With relations (4) and (7), we also ignore the transient changes in mitochondria, such as $[Ca^{2+}]$[8], for sustained thermogenesis such that the following applies

$$dQ = pmf \cdot dH^+ \quad (8)$$

This equation matches the fact that the co-stimulation of neurotransmitters norepinephrine (NE) and ATP can effectively convert the electrochemical potential energy stored in the mitochondrial proton gradient into heat via the mitochondrial uncoupling protein-1 (UCP1) in BA[8].

According to Fourier's law, the relation between heat flux (*J*, heat per unit time per unit area, J s$^{-1}$ m$^{-2}$) and temperature gradient ($\nabla T$, K m$^{-1}$) is written as follows:

$$J = -\kappa \nabla T \qquad (9)$$

Fourier's law is also stated as follows:

$$dQ = JAdt = -\kappa A \nabla T dt \qquad (10)$$

where *dt* is the time interval and *A* is the area. Equations (10) and (8) together give the following

$$\nabla T = -\frac{pmf \cdot dH^+}{\kappa A dt} \qquad (11)$$

We can consider that a mitochondrion with a spherical shape and radius (*r*) has an area $A = 4\pi r^2$, and $dH^+/dt$ is clearly the proton current ($I_{H^+}$) of the mitochondrion. The thermogenic proton current is directed inward and is mediated by UCP1 ($I_{UCP1}$) after its activation. These statements mean that we can rewrite the gradient expression (11) for BA thermogenesis as follows

$$\nabla T = -\frac{pmf}{4\pi \kappa r^2} I_{H^+} = -\frac{pmf}{4\pi \kappa r^2} I_{UCP1} \qquad (12)$$

After determining the equation for the temperature gradient, we can deduce the relation between temperature and time by applying the thermal diffusion equation with a heat source[9]:

$$\frac{\partial T}{\partial t} = D \nabla^2 T + \frac{H}{C} \qquad (13)$$

where $D = \kappa/C$ is the thermal diffusivity (m$^2$ s$^{-1}$), $\kappa$ is the thermal conductivity (W m$^{-1}$ K$^{-1}$), *C* is the volumetric heat capacity (J K$^{-1}$ m$^{-3}$), and heat is generated at a rate *H* per unit volume (W m$^{-3}$, $H = P/V$, *P* is the power, and *V* is the volume)

In spherical polar coordinates[9], we write

$$\nabla^2 T = \frac{1}{r^2} \frac{\partial}{\partial r} \left( r^2 \frac{\partial T}{\partial r} \right) \qquad (14)^9$$

$$\frac{\partial T}{\partial r} = \nabla T \qquad (15)$$

Because $pmf$ and $I_{H^+}$ are not functions of radius ($r$) for a single mitochondrion, equations (14) and (15) together with equation (12) state that the thermogenesis of a mitochondrion satisfies the Laplace equation

$$\nabla^2 T = 0 \qquad (16)$$

Thus, the thermal diffusion equation (13) for a spherical mitochondrion reduces to the following

$$\frac{\partial T}{\partial t} = \frac{H}{C} = \frac{P}{VC} \qquad (17)$$

Dividing both sides of equation (8) by a $dt$ time, we write

$$P = \frac{dQ}{dt} = pmf \cdot \frac{dH^+}{dt} = pmf \cdot I_{H^+} \qquad (18)$$

Equations (17) and (18) yield that the following

$$\frac{\partial T}{\partial t} = \frac{pmf \cdot I_{H^+}}{VC} \qquad (19)$$

In the resting state of BA, without stimulation of sympathetic transmitters, UCP1 is inactivated by purine nucleotides. The BA or mitochondrion has a steady state described according to equation (19) as follows

$$I_{H^+} = I_{UCP1} = 0 \qquad (20)$$

$$\frac{\partial T}{\partial t} = 0 \qquad (21)$$

In the thermogenic state, it is clear that the proton current is not zero and is mediated by the activated UCP1 such that equation (19) states the following

$$I_{H^+} = I_{UCP1} \neq 0 \qquad (22)$$

$$\frac{\partial T}{\partial t} \neq 0 \qquad (23)$$

Using the steady state to discuss the thermogenic state leads to an ~$10^{-5}$ gap between Baffou's model and well-known facts[1]. In our previous paper[7], we noted Baffou's mistakes and properly applied equation (17) for theoretical estimation of the maximum rate of mitochondrial temperature change. The theoretical estimation matched well with the experimental result[7].

After constructing the thermogenic model as a function of time (equation 19), we can further discuss the thermogenic characteristics of BA, such as the thermogenic capacity of the mitochondrion and the limiting factors for BA thermogenesis.

To estimate the temperature profiles of mitochondria, we must know $pmf \cdot I_{H^+}$ in equation (19). Mitchell's chemiosmotic theory states the following

$$pmf = \Delta\psi - 2.3RT/F \cdot \Delta pH \qquad (24)$$

where $\Delta\psi$ is the electrical gradient, $\Delta pH$ is the proton gradient, $R$ is the gas constant, $T$ is the temperature in Kelvin, and $F$ is the Faraday constant. The mitochondrial $pmf$ is ~200 mV. For a single mitochondrion of BA under thermogenesis, the inward thermogenic proton current is the current of the mitoplast, which is mediated by UCP1 ($I_{UCP1}$). It is known that mitoplasts typically have membrane capacitances of 0.5-1.2 pF and proton current ($I_{UCP1}$) densities of 60-110 pA/pF[10].

If defining the change rate of mitochondrial temperature ($\frac{\partial T}{\partial t}$) as a measurement of the thermogenic capacitance in BA, by taking the proton current of mitochondrion as 100 pA and the mitochondrial volume as 1 μm³, we obtain a theoretical rate of mitochondrial $\frac{\partial T}{\partial t}$ of ~4.8 K s⁻¹ based on equation (19).

The maximum experimental thermogenic capacitance of BA is comparable to 10 μM CCCP-induced thermogenesis[8]. However, the measured maximum rate[7] of mitochondrial $\frac{\partial T}{\partial t}$ is ~0.06 K s⁻¹, which suggests that the proton current ($I_{H^+}$) is a limiting factor for BA thermogenesis. For the maximum transient rate of mitochondrial $\frac{\partial T}{\partial t}$, an initial transient [$Ca^{2+}$] change in mitochondria evoked by stimulation of sympathetic transmitters[8] should be counted (equations 6 and 7), which also makes a comparable contribution.

A proton current of 100 pA means that a single mitochondrion consumes $6.24 \times 10^8$ protons per second and that a single BA with ~1000 mitochondria requires $6.24 \times 10^{11}$ proton (~1 pmol) per second. Clearly, free cellular protons are the major limiting factor for thermogenesis, which was experimentally supported by the cytosol alkylation during BA thermogenesis[8].

In equation (3), the change in cellular internal energy is claimed to be negligible. For verification, we calculated the numbers of free available protons, which are $~6.3 \times 10^2$ in a mitochondrion and $~10^5$ in a BA with a diameter of 20 μm and a cytosol pH of 7.4, Thus, we indeed confirmed that $dU$ can be neglected for sustained thermogenesis. Additionally, $dU \approx 0$ suggests that the increased mitochondrial or cellular temperatures must be balanced and compensated by selected intra-mitochondrial or intracellular energy changes, such as exergonic reactions of NADH (52.6 kcal mol$^{-1}$) and FADH$_2$ (43.4 kcal mol$^{-1}$), which were also experimentally supported by NADH and FADH$_2$ consumption during BA thermogenesis[8].

Consequently, the gap between the maximum experimental $\frac{\partial T}{\partial t}$ and the theoretical $\frac{\partial T}{\partial t}$ suggests that thermogenesis of BA uses less than 1% of its thermogenic capacity. In addition, as illustrated in Figure 1, the results demonstrated that the overall averaged $\frac{\partial T}{\partial t}$ was less than ~0.005 K s$^{-1}$. In reality, a single BA might only consume $~10^{-3}$-$10^{-2}$ pmol proton per second for sustained thermogenesis in BA (Figure 1). Furthermore, depolarization of the mitochondrial membrane potential and cytosol alkylation during BA thermogenesis[8] suggest that the value of $pmf \cdot I_{H^+}$ is a factor of self-restriction for thermogenesis.

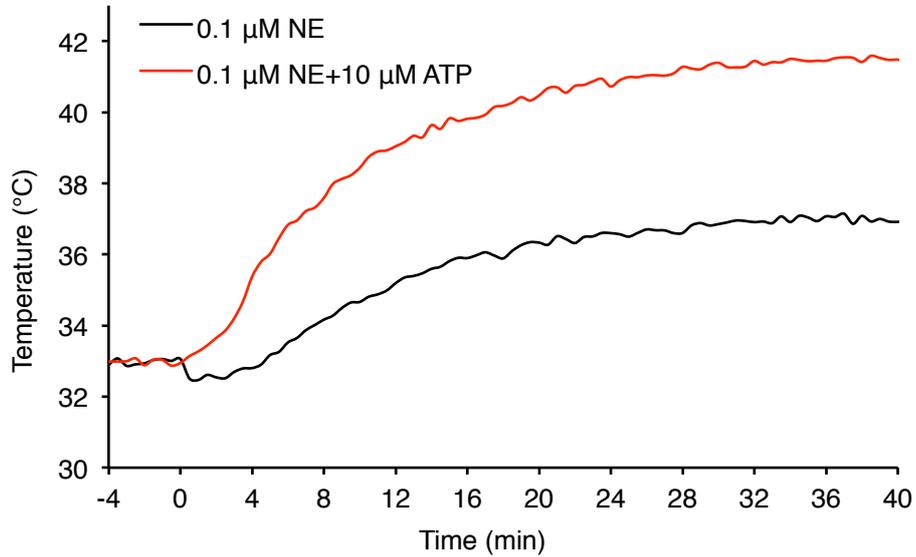

**Figure 1** Change profiles of mitochondrial temperature in BA under stimulation starting from 0 min of 0.1 μM NE without (black line) or with (red line) 10 μM ATP-induced thermogenesis in BA[8].

In summary, BA and its mitochondria are heat-producing micro-machines with a high efficacy limited by free proton pools. The thermogenic model (equation 19) and calculations suggest that BA thermogenesis relies on hydrogen and energy sources such as glucose, water, fatty acid, NADH, and FADH$_2$. One mol glucose and 6 mol water together can supply 24 mol protons in the tricarboxylic acid cycle. Even if glucose is supplied at a rate[11] of 0.18 pmol h$^{-1}$ cell$^{-1}$ without tens or hundreds of times the glucose uptake in BA under stimulation[12,13], it is sufficient to sustain thermogenesis in BA.

$$\frac{\partial T}{\partial t} = \eta \frac{pmf \cdot I_{H^+}}{VC} \quad (25)$$

Finally, the thermogenic model of mitochondria (equation 19) can be generalized as shown in equation (25) by multiplying the thermogenic efficiency ($\eta$). The thermogenesis of BA under NE and ATP co-stimulation is a special case ($\eta = 1$) for equation (25). In general, of the potential free energy in glucose, approximately 40 percent is conserved in ATP in mitochondrial oxidative phosphorylation, and thus, the value of $\eta$ is approximately 0.6 for all other cell types except erythrocytes ($\eta = 0$), which lack mitochondria. Interestingly, NE stimulation alone activates the proton-pumping ATPase function of mitochondrial complex V in BA[8], so that NE-stimulated BA show a variety of responses (heating, constant temperature or occasionally cooling) and a low efficacy of thermogenesis[8]. Consequently, the phenomena and equation (25) restate that the limiting factor for the capability of intracellular thermogenesis is the net proton current ($I_{H^+}$) of proton outflow by proton pumps and proton leakage by UCP1 or other uncoupling factors.

**References**

1. Baffou, G., Rigneault, H., Marguet, D. & Jullien, L. A critique of methods for temperature imaging in single cells. *Nat. Methods* **11,** 899–901 (2014).

2. Kiyonaka, S. *et al.* Validating subcellular thermal changes revealed by fluorescent thermosensors. *Nat. Methods* **12,** 801–802 (2015).

3. Suzuki, M., Zeeb, V., Arai, S., Oyama, K. & Ishiwata, S. The 105 gap issue between calculation and measurement in single-cell thermometry. *Nat. Methods* **12,** 802–803 (2015).

4. Baffou, G., Rigneault, H., Marguet, D. & Jullien, L. Reply to: 'Validating subcellular thermal changes revealed by fluorescent thermosensors' and 'The 105 gap issue between calculation and measurement in single-cell thermometry'. *Nat. Methods* **12,** 803–803 (2015).

5. Chrétien, D. *et al.* Mitochondria are physiologically maintained at close to 50 °C. *PLOS Biol.* **16,**

e2003992 (2018).

6. Lane, N. Hot mitochondria? *PLOS Biol.* **16,** e2005113 (2018).

7. Xie, T.-R., Liu, C.-F. & Kang, J.-S. Dye-based mito-thermometry and its application in thermogenesis of brown adipocytes. *Biophys. Rep.* **3,** 85–91 (2017).

8. Xie, T.-R., Liu, C.-F. & Kang, J.-S. Sympathetic transmitters control thermogenic efficacy of brown adipocytes by modulating mitochondrial complex V. *Signal Transduct. Target. Ther.* **2,** sigtrans201760 (2017).

9. Blundell, S. & Blundell, K. M. *Concepts in Thermal Physics*. (OUP Oxford, 2010).

10. Bertholet, A. M. *et al.* Mitochondrial Patch Clamp of Beige Adipocytes Reveals UCP1-Positive and UCP1-Negative Cells Both Exhibiting Futile Creatine Cycling. *Cell Metab.* **25,** 811–822.e4 (2017).

11. Zamorano, F., Wouwer, A. V. & Bastin, G. A detailed metabolic flux analysis of an underdetermined network of CHO cells. *J. Biotechnol.* **150,** 497–508 (2010).

12. Orava, J. *et al.* Different Metabolic Responses of Human Brown Adipose Tissue to Activation by Cold and Insulin. *Cell Metab.* **14,** 272–279 (2011).

13. Vallerand, A. L., Perusse, F. & Bukowiecki, L. J. Stimulatory effects of cold exposure and cold acclimation on glucose uptake in rat peripheral tissues. *Am. J. Physiol. - Regul. Integr. Comp. Physiol.* **259,** R1043–R1049 (1990).

**Acknowledgements** The author thanks Dr. Xiao-Feng Liu and Dr. Tao-Rong Xie for discussions.

**Competing financial interests**

The authors declare no competing financial interests.

**Methods**

The thermogenic model (equation 19) overcomes obstacles related to the ability to increase cellular temperature by endogenous thermogenesis. Therefore, quantification is needed, which has not been performed in our previous works. Thus, the temperatures were calculated and converted from our previous data[8,7]. The calculation was based on the relation (equation 25) between temperature ($T$) and the normalized intensity ratio ($nr$) of thermosensitive and thermoneutral mitochondrial dyes[7].

$$\frac{1}{T} - \frac{1}{T_{ref}} = -\frac{k_B}{E_a} \cdot \ln nr \qquad (25)$$

where $k_B$ is the Boltzmann constant, and $E_a$ is the measured activation energy (~6.55 kcal/mol)[7].